\documentclass{ws-mpla}
\usepackage[super]{cite}
\usepackage{graphicx}
\RequirePackage{lineno}

\newcommand{\snn}{$\sqrt{s_{\mathrm{NN}}}$~}

\newcommand{\s}{$\sqrt{s}$~}

\newcommand{\pb}{\mbox{Pb--Pb}~}
\newcommand{\rh}{\ensuremath{\mathrm{\rho_{00}}}~}
\newcommand{\rha}{\ensuremath{\mathrm{\rho_{00}}}}
\newcommand{\ct}{$\cos{\theta^{*}}$~}

\newcommand{\ksta}{$K^{*0}$}
\newcommand{\kst}{$K^{*0}$~}
\newcommand{\akst}{$\overline{K}^{*0}$~}

\newcommand{\kzs}{\mbox{$K^0_S$}~}

\newcommand{\pha}{$\phi$}
\newcommand{\ph}{$\phi$~}
\newcommand{\pt}{$p_{\mathrm{T}}$~}
\newcommand{\pta}{$p_{\mathrm{T}}$}

\begin{document}

\markboth{Bedangadas Mohanty, Sourav Kundu, Subhash Singha and Ranbir Singh}
{Spin alignment measurement of vector mesons produced in high energy collisions}

%%%%%%%%%%%%%%%%%%%%% Publisher's Area please ignore %%%%%%%%%%%%%%
\catchline{}{}{}{}{}
%%%%%%%%%%%%%%%%%%%%%%%%%%%%%%%%%%%%%%%%%%%%%%%%%%%%%%%%%%%%%%%%%%%
\title{Spin alignment measurement of vector mesons produced in high energy collisions}
\author{Bedangadas Mohanty}
\address{School of Physical Sciences, National Institute of Science Education and Research, HBNI,\\ Jatni-752050, India\\
bedanga@niser.ac.in}
\author{Sourav Kundu}
\address{Experimental Physics Department, CERN, CH-1211 Geneva 23, Switzerland}
\author{Subhash Singha}
\address{Institute of Modern Physics, Chinese Academy of Sciences Lanzhou, \\Gansu 730000, China}
\author{Ranbir Singh}
\address{School of Physical Sciences, National Institute of Science Education and Research, HBNI,\\  Jatni-752050, India}
\maketitle
\pub{Received (Day Month Year)}{Revised (Day Month Year)}
\begin{abstract}
This review covers the recent experimental development on spin alignment measurements of \kst and \ph vector mesons in heavy-ion and pp collisions at RHIC and LHC energies. Measurements in $e^{+}e^{-}$ collisions at LEP energies are also discussed. Spin alignment of vector mesons are studied by measuring the second diagonal element \rh of spin density matrix. The spin density matrix element \rh is obtained by measuring the angular distribution of vector meson decay daughter with respect to the quantization axis in vector meson rest frame. Measured \rh values for vector mesons are found to be larger than 1/3 at high momentum in $e^{+}e^{-}$ collisions at LEP energies, suggesting the preferential production of vector meson with helicity zero state from the fragmentation process. The \rh values are found to be smaller than 1/3 (\rh = 1/3 implies no spin alignment) for \kst and \ph vector mesons at low transverse momentum in Pb--Pb collisions at \snn = 2.76 TeV. This observations are qualitatively consistent with the expectation from models which attribute the spin alignment effect due to polarization of quarks in the presence of large initial angular momentum in non-central heavy-ion collisions and its subsequent hadronization by the process of recombination. No significant spin alignment effect is observed for \kzs (spin = 0) in mid-central Pb--Pb collisions and for vector mesons in pp collisions. 
However, the preliminary results of \rh for \ph mesons are larger than 1/3 at intermediate \pt in Au--Au collisions at RHIC energies and can be attributed to the presence of \ph meson field. Although there is evidence of spin alignment effect of vector mesons in heavy-ion collisions but the measured effect is surprisingly larger in context of hyperon polarization. Therefore these results will trigger further theoretical study. 

\end{abstract}
\ccode{PACS Nos.: include PACS Nos}
%\ccode{25.75.−q,25.75.Nq., 13.88.+e, 24.70.+s, 13.66.Bc, 3.75.Cs}

\section{Introduction}	
The spin-orbital angular momentum interactions are one of the most important effect in nuclear, atomic and condensed matter physics. This causes fine structure and shell structure in atomic and nuclear physics, respectively and is a key ingredient in the field of spintronics in material sciences. The spin-orbital angular momentum interaction is also expected to affect the evolution of hot and dense quark-gluon plasma (QGP), produced in ultra-relativistic heavy-ion collisions. High energy heavy-ion collisions create a temperature more than 10$^{13}$ K at RHIC and 10$^{16}$ K at LHC energies. Under this extreme condition quarks and gluons inside protons and neutrons are set free over an extended volume and leads to the formation of QGP. Heavy-ion collisions offer an ideal environment for studying the properties of QGP phase and test various phenomenon of quantum chromodynamics (QCD)~\cite{Braun-Munzinger:2007edi}.

Given the importance of spin-orbit interactions in several fields of physics, it is imperative to look for its possible effect on particles with non-zero spin in a system with high orbital angular momentum. In non-central heavy-ion collisions, where the impact parameter ($b$) between two colliding nuclei is non-zero, a very large orbital angular momentum (OAM) of the order of 10$^{5}$ -- 10$^{7}$ $\hslash$ is expected to be created about the centroid of the participant matter~\cite{sa1}. In presence of such large OAM, spin-orbit coupling of QCD could lead to the polarization of quarks and anti-quarks in the produced QGP medium. The polarization of quarks and anti-quarks are further translated to the polarization of produced particles with non-zero spin along the direction of the OAM during the process of hadronization~\cite{sam2,sam3,sam4}. The transfer of OAM to the QGP medium in the form of preferential alignment of the spin of the produced particles along the OAM direction is known as global polarization. Statistical mechanics, kinetic theory and hydrodynamics further show that the OAM can be manifested in the form of fluid vorticity~\cite{Li:2017slc, Florkowski:2017ruc, Becattini:2017gcx, Weickgenannt:2019dks}.

There are two kinds of polarization of produced hadrons which have been studied in experiment: 1) vector polarization and 2) tensor polarization. The vector polarization has been studied by measuring the polarization of hyperons, whereas the tensor polarization is studied by measuring the polarization of vector mesons. Unlike the polarization of hyperons, which undergo weak decay with parity violation, the polarization of vector meson can not be measured directly as they mainly decay through the parity conserving strong decay. However, the spin alignment of vector mesons can be studied by measuring the diagonal elements of 3$\times$3 hermitian spin density matrix with unit trace. The diagonal elements $\rho_{11}$, $\rho_{00}$ and $\rho_{-1-1}$ are the probabilities of the spin component of vector mesons along the quantization axis. Among these three diagonal elements, $\rho_{00}$ is independent, whereas $\rho_{11}$ and $\rho_{-1-1}$ cannot be measured separately in two-body decays to pseudoscalar mesons. Therefore, in experiment, spin alignment of vector mesons has been studied by measuring the \rh element of the spin density matrix. In the absence of spin alignment, all the three spin states of vector mesons 1, 0 and -1 are equally probable and that makes \rh = 1/3. Any deviation of \rh from 1/3 is the experimental signature of the spin alignment of the vector mesons. In high energy experiment the diagonal element \rh is measured from the angular distribution of the vector meson decay daughter with respect to a quantization axis, in vector meson rest frame. Unlike hyperon polarization, spin alignment of vector meson has negligible contamination from the decays as they are predominantly produced directly. For the cases of two body strong decay of vector mesons there is no uncertainty due to the decay parameter as seen for hyperons.

Along with the distinct importance of spin alignment study of vector mesons in heavy-ion collisions, these measurements are also carried out in high energy lepton and hadron collisions in order to understand the production mechanism of vector mesons. In high energy collisions, spin alignment of vector mesons can occur from the fragmentation of un-polarized quarks. Therefore, these measurements are useful to understand the spin dependent fragmentation function and hence considered as one of the important aspects in high energy spin physics.

In this paper, we review the experimental results of the spin alignment of vector mesons in high energy heavy-ion, hadron and lepton collisions. This review is organized as follows: in next section we discuss the experimental observable, followed by a brief description about the coordinate systems used in the experimental analysis. In section~\ref{analysis} we briefly discuss the analysis techniques. In section~\ref{results_pp} and section~\ref{results_AA} we review the experimental results of spin alignment of vector mesons in $e^{+}e^{-}$ and hadron-hadron collisions, and in heavy-ion collisions, respectively. A brief theoretical discussion related to the experimental results are also presented in section~\ref{results_pp} and section~\ref{results_AA}. Finally a summary and future prospects are presented in section~\ref{summary}

\section{Experimental observable: angular distribution of vector meson's decay daughter}
The spin alignment of vector mesons are studied by measuring the diagonal element $\rho_{00}$ of a 3$\times$3 hermitian spin density matrix~\cite{schilling}. $\rho_{00}$ corresponds to the probability of finding a vector meson in spin state 0 out of 3 possible spin states of -1, 0 and 1. All three states are equally probable in absence of spin alignment. This leads to $\rho_ {00}$ = 1/3. On the other hand, in the presence of spin alignment $\rho_{00}$ will deviate from 1/3. In experiment, the spin density matrix element $\rho_{00}$ is measured by studying the angular distribution of the decay daughter of vector meson with respect to a quantization axis in vector meson's rest frame. In order to obtain the angular distribution of vector meson decay daughter, let us consider a vector meson is at rest and decays to two spin 0 particles. The projection of the total angular momentum ($\vec{J}$ = 1) of vector meson along any arbitrary quantization axis is $m$. The completeness relation of vector meson state gives,
\begin{equation}
\sum_{m~=~-1,~0,~1}|1,~m\rangle \langle 1,~m| = 1. 
\label{eq1}
\end{equation} 
In the rest frame of vector meson, momentum of two decay daughters are back to back and is denoted as $\vec{p}(\theta^{*}, \phi^{*})$, where $\theta^{*}$ is the polar angle made by vector meson's decay daughter with the quantization axis and $\phi^{*}$ is the corresponding azimuthal angle. Using the conservation of final and initial state total angular momentum, two particle final state of decay daughters of vector meson can be expressed in terms of their helicities ($\lambda_{1}$, $\lambda_{2}$), total angular momentum, polar and azimuthal angle as,
\begin{equation}
|\theta^{*},~\phi^{*},~J=1,~\lambda\rangle,
\end{equation}
where $J$ corresponds to the total angular momentum and the difference between the helicities of two decay daughters is $\lambda$. In helicity basis, the final state physically corresponds to a state which have total angular momentum 1 and its projection $\lambda$ along the flight direction of any one of the decay daughter which is expressed in terms of $\theta^{*}$ and $\phi^{*}$. The $\lambda$ is given as,
\begin{equation}
\lambda~ =~\lambda_{1}~-~\lambda_{2}~=~(\vec{s_{1}}~-~\vec{s_{2}}).\hat{p},
\end{equation}
where $s_{1}$ and $s_{2}$ are the spin of the decay daughters. As the both decay daughters are spin 0 particle, that makes $\lambda$ = 0. Therefore the two particle final state of vector meson's decay daughters can be expressed as $|\theta^{*}, \phi^{*}, 1, 0\rangle$.

The angular distribution of vector meson decay daughters in the rest frame of vector meson can be expressed as,
\begin{equation}
\frac{\mathrm{d}N}{\mathrm{d}cos\theta^{*}\mathrm{d}\phi^{*}}~=~\langle \theta^{*},~\phi^{*},~1,~0|T \rho T^{\dag}|\theta^{*},~\phi^{*},~1,~0\rangle,
\label{eq2}
\end{equation} 
where $T$ is the transition matrix and $\rho$ is the spin density matrix. By substituting Eq.~(\ref{eq1}) in Eq.~(\ref{eq2}), the decay angular distribution can be expressed as,
\begin{eqnarray}
\frac{\mathrm{d}N}{\mathrm{d}cos\theta^{*}\mathrm{d}\phi^{*}}~&=&~\sum_{m}\sum_{m^{\prime}}~\langle \theta^{*},~\phi^{*},~1,~0|T|1,~M\rangle \nonumber \\
&& \langle 1,~m| \rho |1,~m^{\prime}\rangle \langle 1,~m^{\prime}| T^{\dag}|\theta^{*},~\phi^{*},~1,~0\rangle.
\label{eq3}
\end{eqnarray}
where $\langle\theta^{*}, \phi^{*}, 1, 0|T|1, m\rangle$ is the transition amplitude of a state $|1, 0\rangle$ from the state $|1, m\rangle$. Using the Wigner D-matrix formalism~\cite{rose}, the decay amplitude is written as,
\begin{equation}
\langle\theta^{*},~\phi^{*},~1,~0|T|1,~m\rangle~=~c~D_{m,~0}^{1\dag}~(\phi^{*},~\theta^{*},~-\phi^{*}),
\label{eq4}
\end{equation}
where $D$ is the Wigner $D$-matrix element and $c$ is the normalization constant. By substituting Eq.~(\ref{eq4}) in Eq.~(\ref{eq3}), we get
\begin{equation}
\frac{\mathrm{d}N}{\mathrm{d}cos\theta^{*}\mathrm{d}\phi^{*}}~=~~\sum_{T}\sum_{T^{\prime}}~D_{m,~0}^{1\dag}~\rho_{m,~m^{\prime}}~D_{m^{\prime},~0}^{1},
\label{eq5}
\end{equation}
where $\rho_{m, m^{\prime}}$ = $\langle1, m|\rho|1, m^{\prime}\rangle$ are the density matrix elements. The Wigner $D$-matrix elements are given by,
\begin{equation}
D_{1,~0}^{1}~=~-\frac{1}{2}~\sin\theta^{*}~\exp^{-i\phi^{*}},
\label{eq6}
\end{equation}
\begin{equation}
D_{0,~0}^{1}~=~-\cos\theta^{*},
\label{eq7}
\end{equation}
\begin{equation}
D_{-1,~0}^{1}~=~\frac{1}{2}~\sin\theta^{*}~\exp^{i\phi^{*}}.
\label{eq8}
\end{equation}

Using the Wigner $D$-matrix elements, Eq.~(\ref{eq5}) can be expressed as,
\begin{eqnarray}
\frac{\mathrm{d}N}{\mathrm{d}\cos{\theta^{*}}\mathrm{d}\phi^{*}}  &\propto& [\cos^{2}{\theta^{*}}\rho_{00} + \sin^{2}{\theta^{*}}(\rho_{11}+\rho_{-1-1})/2 \nonumber \\
&&~+~\sin{2\theta^{*}}\cos{\phi^{*}}(\mathrm{Re}\rho_{-10} - \mathrm{Re}\rho_{01})/\sqrt{2} \nonumber \\
&&~+~\sin{2\theta^{*}}\sin{\phi^{*}}(\mathrm{Im}\rho_{-10} - \mathrm{Im}\rho_{01})/\sqrt{2} \nonumber \\
&&~-~\sin^{2}{\theta^{*}}(\cos{2\phi^{*}}\mathrm{Re}\rho_{1-1} + \sin{2\phi^{*}}\mathrm{Im}\rho_{-11})]
\label{eq9}
\end{eqnarray}
The unit trace condition of spin density matrix elements $\rho_{-1-1} + \rho_{00} + \rho_{11}~=~1$ and integration over azimuthal angle leads Eq.~(\ref{eq9}) to
\begin{eqnarray}
 \frac{\mathrm{d}N}{\mathrm{d}cos\theta^{*}}~\propto~[1~-~\rho_{00}~+~(3\rho_{00}~-~1)cos^{2}\theta^{*}].
 \label{ang_dist}
\end{eqnarray}

\section{Coordinate system}
\begin{figure}[h]
\centering 
\includegraphics[scale=0.5]{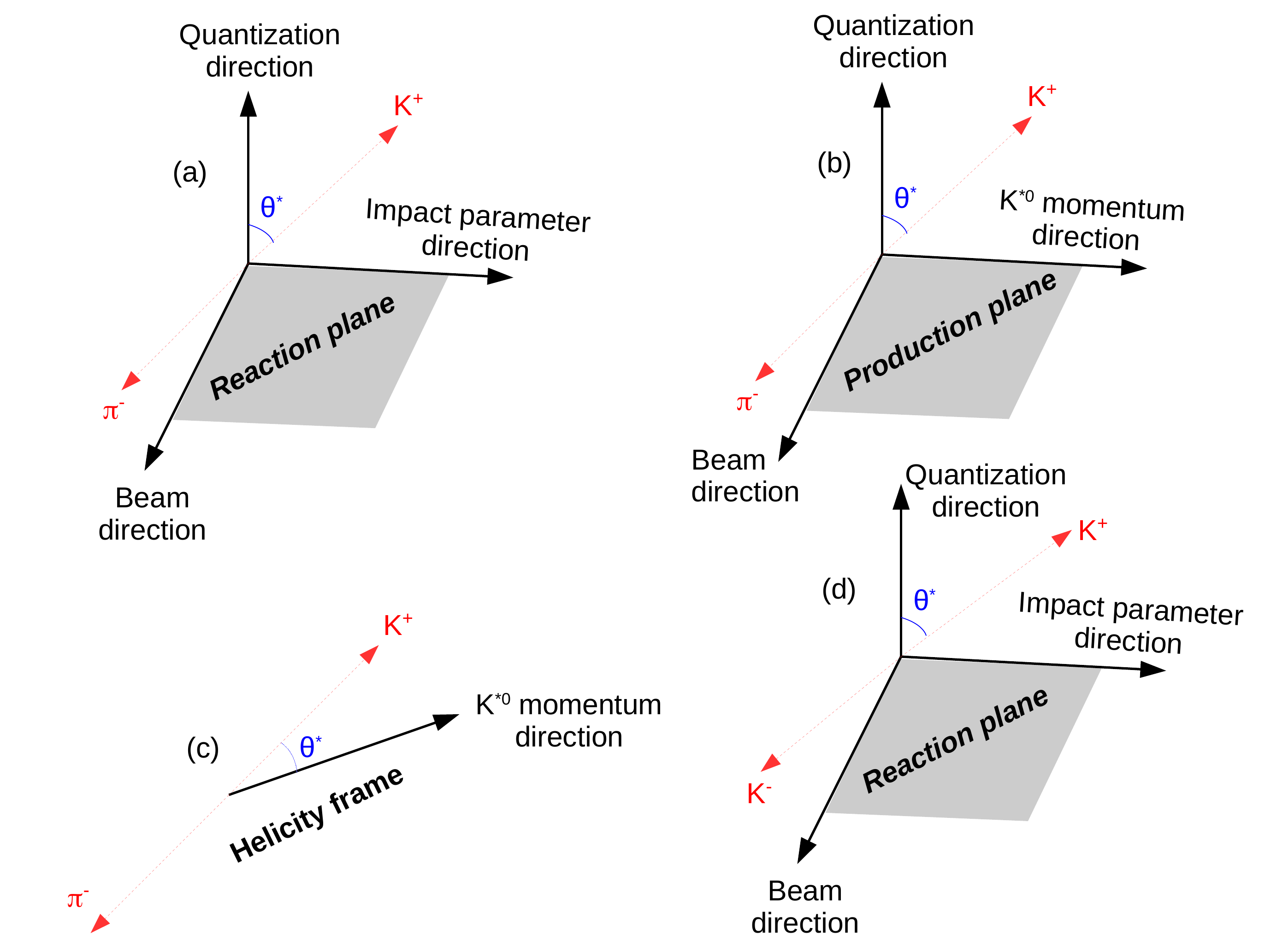}
\caption{A graphical illustration of various frame of reference used in spin alignment measurement of vector meson.}
\label{coordinate} 
\vspace{-3mm}
\end{figure}

\begin{figure}[h]
\begin{center}
\includegraphics[scale=0.5]{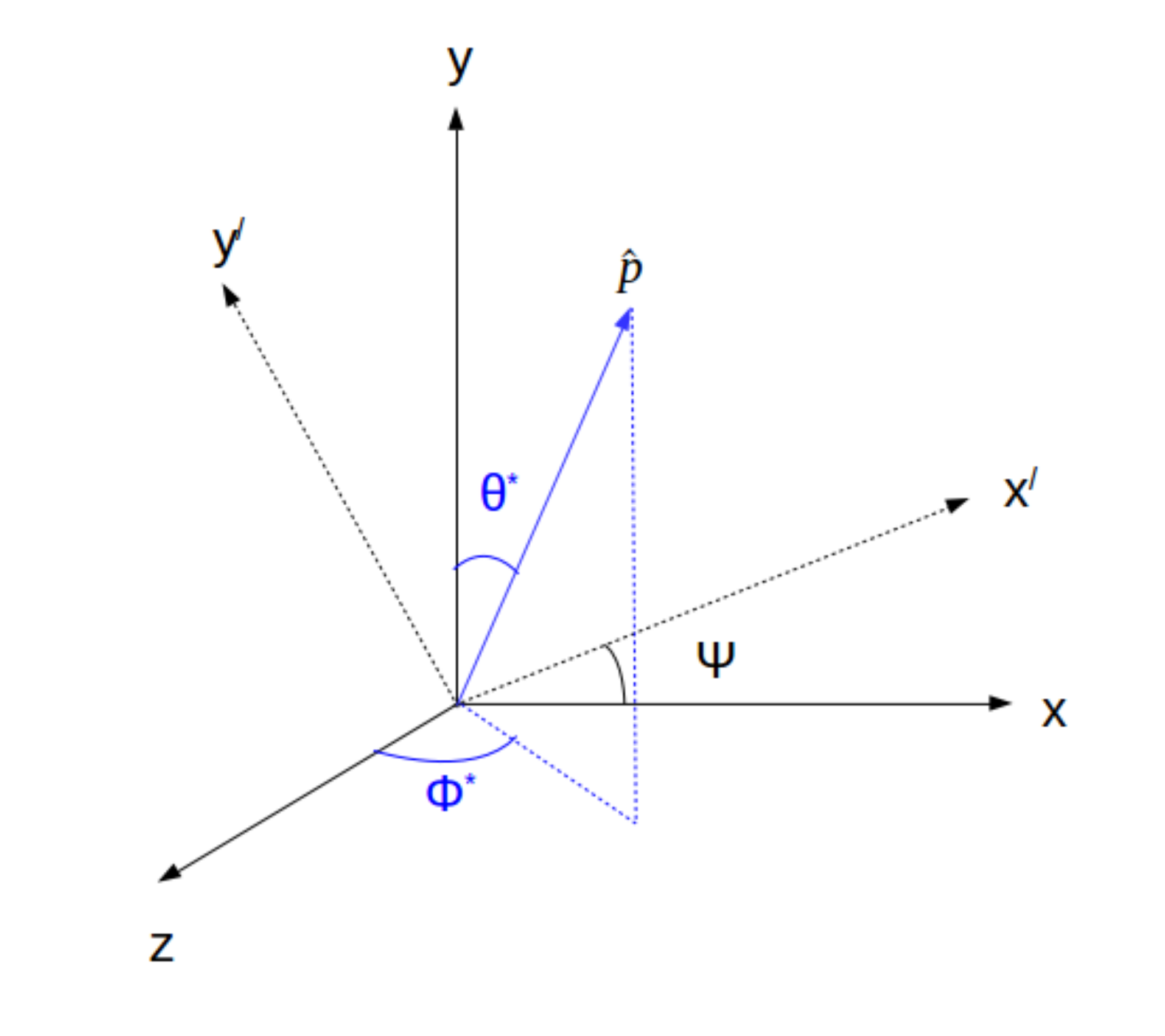}
\caption{The reaction plane coordinate system along with the event plane vector obtained by rotating $xy$ plane by angle $\psi$ with respect to $z$ axis.}
\label{fig2}
\end{center}
\vspace{-3mm}
\end{figure}

In experiment, three widely used coordinate system for spin alignment measurements are 1) helicity frame, 2) production plane and 3) reaction plane. In helicity frame the quantization axis is the momentum direction of the vector meson in lab frame and $\theta^{*}$ is the angle made by vector meson decay daughter with the quantization axis in vector meson rest frame as shown in the panel (c) of Fig.~\ref{coordinate}. The panel (b) of Fig.~\ref{coordinate} shows the graphical illustration of the production plane. In production plane analysis quantization axis is the perpendicular direction of the production plane which is defined by the beam direction and the momentum direction of the vector meson. The reaction plane is defined by the beam direction and the impact parameter direction, and perpendicular direction to the reaction plane can also be used as a quantization axis as shown in the panel (a) and (d) of Fig.~\ref{coordinate} for \kst and \ph, respectively. The perpendicular direction of the reaction plane also corresponds the direction of the initial global angular momentum. 
In experiment, we can not measure the impact parameter direction. Therefore, we use the event plane as a proxy of the reaction plane and further correct the measurements with event plane resolution. The event plane is estimated from the azimuthal angle of the produced hadrons and can be expressed in terms of event plane vector as,
\begin{equation}
\vec{Q}~=\frac{1}{n}\big[\sum_{i=1}^{N}\sin{n\phi_{i}}~\hat{i}~+~\sum_{i=1}^{N}\cos{n\phi_{i}}~\hat{j}\big],
\end{equation} 
where the $\vec{Q}$ is the event plane vector, $n$ is the order of the event plane, summation is over all produced charged particle and $\phi_{i}$ is the azimuthal angle of the $i^{\mathrm{th}}$ charged particle. In order to derive the relation between measured $\rho_{00}$ w.r.t. the event plane and the reaction plane, let us assume a right handed coordinate system with the impact parameter direction, the beam direction and the angular momentum direction are along $x$, $z$ and $y$ axis, respectively. The reaction plane is represented by the $xz$ plane. In the rest frame of vector meson the unit momentum vector ($\hat{p}$) of one of its decay daughter makes the polar angle  $\theta^{*}$ and the azimuthal angle $\phi^{*}$. Let us consider the event plane vector is along $\mathrm{x}^{\prime}$ and can be obtained by a rotation of angle $\psi$ about the $\mathrm{z}$ axis. Figure~\ref{fig2} shows the reaction plane coordinate system along with the event plane vector.
The distribution of the event plane vector is centered around the $x$ with finite resolution. The event plane resolution ($R$) can be estimated by averaging $\psi$ over a large number of events. The distribution of $\psi$ over many events is an even function centered at zero that gives
\begin{eqnarray}
R~=~\langle~\cos(2\psi)~\rangle,~\mathrm{and}~\langle~\sin(2\psi)~\rangle~=~0.
\end{eqnarray}
The angular distribution of decay daughter of vector meson with respect to the quantization axis perpendicular to the reaction plane can be expressed as,
\begin{eqnarray}
\frac{\mathrm{d}N}{\mathrm{d}\cos{\theta^{*}}\mathrm{d}\phi^{*}}  &\propto& [\cos^{2}{\theta^{*}}\rho^{\mathrm{RP}}_{00} + \sin^{2}{\theta^{*}}(\rho^{\mathrm{RP}}_{11}+\rho_{-1-1})/2 \nonumber \\
&&~+~\sin{2\theta^{*}}\cos{\phi^{*}}(\mathrm{Re}\rho^{\mathrm{RP}}_{-10} - \mathrm{Re}\rho^{\mathrm{RP}}_{01})/\sqrt{2} \nonumber \\
&&~+~\sin{2\theta^{*}}\sin{\phi^{*}}(\mathrm{Im}\rho^{\mathrm{RP}}_{-10} - \mathrm{Im}\rho^{\mathrm{RP}}_{01})/\sqrt{2} \nonumber \\
&&~-~\sin^{2}{\theta^{*}}(\cos{2\phi^{*}}\mathrm{Re}\rho^{\mathrm{RP}}_{1-1} + \sin{2\phi^{*}}\mathrm{Im}\rho^{\mathrm{RP}}_{-11})]
\label{res_eq0}
\end{eqnarray}
The unit vectors in the reaction plane frame and in the event plane frame are related by,
\begin{eqnarray}
\hat{x}~=~\cos(\psi)\hat{x^{\prime}}~-~\sin(\psi)\hat{y^{\prime}},
\label{res_eq1}
\end{eqnarray}
\begin{eqnarray}
\hat{y}~=~\sin(\psi)\hat{x^{\prime}}~+~\cos(\psi)\hat{y^{\prime}},
\label{res_eq2}
\end{eqnarray}
and
\begin{eqnarray}
\hat{z}~=~\hat{z^{\prime}}.
\label{res_eq3}
\end{eqnarray}
Let us consider in event plane frame, the polar and azimuthal angles are $\theta^{\prime}$ and $\phi^{\prime}$. The momentum direction of vector meson's decay daughter in two different frames can be expressed as,
\begin{eqnarray}
\hat{p}~&=&~\sin{\theta^{*}}\cos{\phi^{*}}~\hat{z}~+~\sin{\theta^{*}}\sin{\phi^{*}}~\hat{x}~+~\cos{\theta^{*}}~\hat{y}\nonumber \\
&&~=~\sin{\theta^{*\prime}}\cos{\phi^{*\prime}}~\hat{z^{\prime}}~+~\sin{\theta^{*\prime}}\sin{\phi^{*\prime}}~\hat{x^{\prime}}~+~\cos{\theta^{*\prime}}~\hat{y^{\prime}}.
\label{res_eq4}
\end{eqnarray}
By substituting Eq.~(\ref{res_eq1}),~(\ref{res_eq2}) and ~(\ref{res_eq3}) in Eq.~(\ref{res_eq4}) we get,
\begin{eqnarray}
\hat{p}.\hat{z}~=~\sin{\theta^{*}}\cos{\phi^{*}}~=~\sin{\theta^{*\prime}}\cos{\phi^{*\prime}}.
\label{res_eq6}
\end{eqnarray}
\begin{eqnarray}
\hat{p}.\hat{x}~=~\sin{\theta^{*}}\sin{\phi^{*}}~=~\sin{\theta^{*\prime}}\sin{\phi^{*\prime}}\cos{\psi}~-~\cos{\theta^{*\prime}}\sin{\psi}.
\label{res_eq7}
\end{eqnarray}
\begin{eqnarray}
\hat{p}.\hat{y}~=~\cos{\theta^{*}}~=~\sin{\theta^{*\prime}}\sin{\phi^{*\prime}}\sin{\psi}~+~\cos{\theta^{*\prime}}\cos{\psi}.
\label{res_eq8}
\end{eqnarray}
By substituting Eq.~(\ref{res_eq6}), Eq.~(\ref{res_eq7}) and Eq.~(\ref{res_eq8}) in Eq.~(\ref{res_eq0}) and after integrating over $\phi^{*\prime}$ we get,
\begin{eqnarray}
\frac{\mathrm{d}N}{\mathrm{d}\cos{\theta^{*\prime}}} &\propto& \left[1~-~\{\rho^{\mathrm{RP}}_{00}~-~\frac{1}{2}\sin^{2}{\psi}(3\rho^{\mathrm{RP}}_{00}~-~1)\}\right]\nonumber \\
&&~+~\left[3\{\rho^{\mathrm{RP}}_{00}~-~\frac{1}{2}\sin^{2}{\psi}(3\rho^{\mathrm{RP}}_{00}~-~1)\}~-~1\right]\cos^{2}{\theta^{*\prime}}\nonumber \\
&&~\propto~(1~-~\rho^{EP}_{00})~+~(3\rho^{EP}_{00}~-~1)\cos^{2}{\theta^{*\prime}}.
\label{res_eq9}
\end{eqnarray}
From Eq.~(\ref{res_eq9}), the relation between $\rho^{\mathrm{RP}}_{00}$ and $\rho^{\mathrm{EP}}_{00}$ can be written as,
\begin{eqnarray}
\rho^{\mathrm{EP}}_{00}~=~\rho^{\mathrm{RP}}_{00}~-~\frac{1}{2}\sin^{2}{\psi}(3\rho^{\mathrm{RP}}_{00}~-~1).
\label{res_eq10}
\end{eqnarray}
Event average of  Eq.~\ref{res_eq10} gives, 
\begin{eqnarray}
\rho^{\mathrm{RP}}_{00}~-~\frac{1}{3}=~\left(\rho^{\mathrm{EP}}_{00}~-~\frac{1}{3}\right)\frac{4}{1~+~3R}.
\label{res_eq11}
\end{eqnarray}
%\section{Theory overview}

\section{\label{analysis}Analysis technique}
\kst and \ph are short lived resonance particles and can not be detected directly in detector. Therefore, \kst and \ph vector mesons are reconstructed from the invariant mass distribution of oppositely charged daughter pairs (\kst (\akst) $\rightarrow$ $K^{+}(K^{-})\pi^{-}(\pi^{+})$) and \ph $\rightarrow$ $K^{+}K^{-}$). Charged decay daughters are identified with various particle identification techniques, depending upon the detector configuration. Two such particle identification techniques used in the ALICE and the STAR experiments are: the specific energy loss measured in the Time Projection Chamber (TPC) and the velocity measured by the Time-Of-Flight (TOF) detector. The invariant mass distribution of resonance decay daughters consists of resonance signal and a large combinatorial background.  The combinatorial background can be estimated by different techniques such as mixed event background, like sign background and rotational background. After the combinatorial background subtraction resonance signal is visible on top of a residual background, which comes due to the production of correlated daughter pairs from the decays of other hadrons and from jets. Resonance signal along with the residual background are fitted with combination of a signal fit function and a background fit function. The Breit-Wigner (Voigtian) distribution function is widely used to describes the \kst (\pha) signal, whereas a second order polynomial is used to describe the residual background. Vector mesons signals are extracted in various $\cos{\theta^{*}}$ bins and are corrected for the detector acceptance $\times$ efficiency to get the corrected yields. A Monte Carlo simulation of the detector response based on a GEANT simulation is used to determine the acceptance $\times$ efficiency. The acceptance and efficiency corrected $\cos{\theta^{*}}$ distributions are fitted with Eq.~(\ref{ang_dist}) to extract \rh values for vector mesons. Details about the analysis techniques can be found in~\cite{alice, star2007}. An alternative analytic approach for correcting the acceptance effects in \rh measurement is proposed in.~\cite{analytic_accep_corr}
\begin{figure}[h]
\centering 
\includegraphics[scale=0.3]{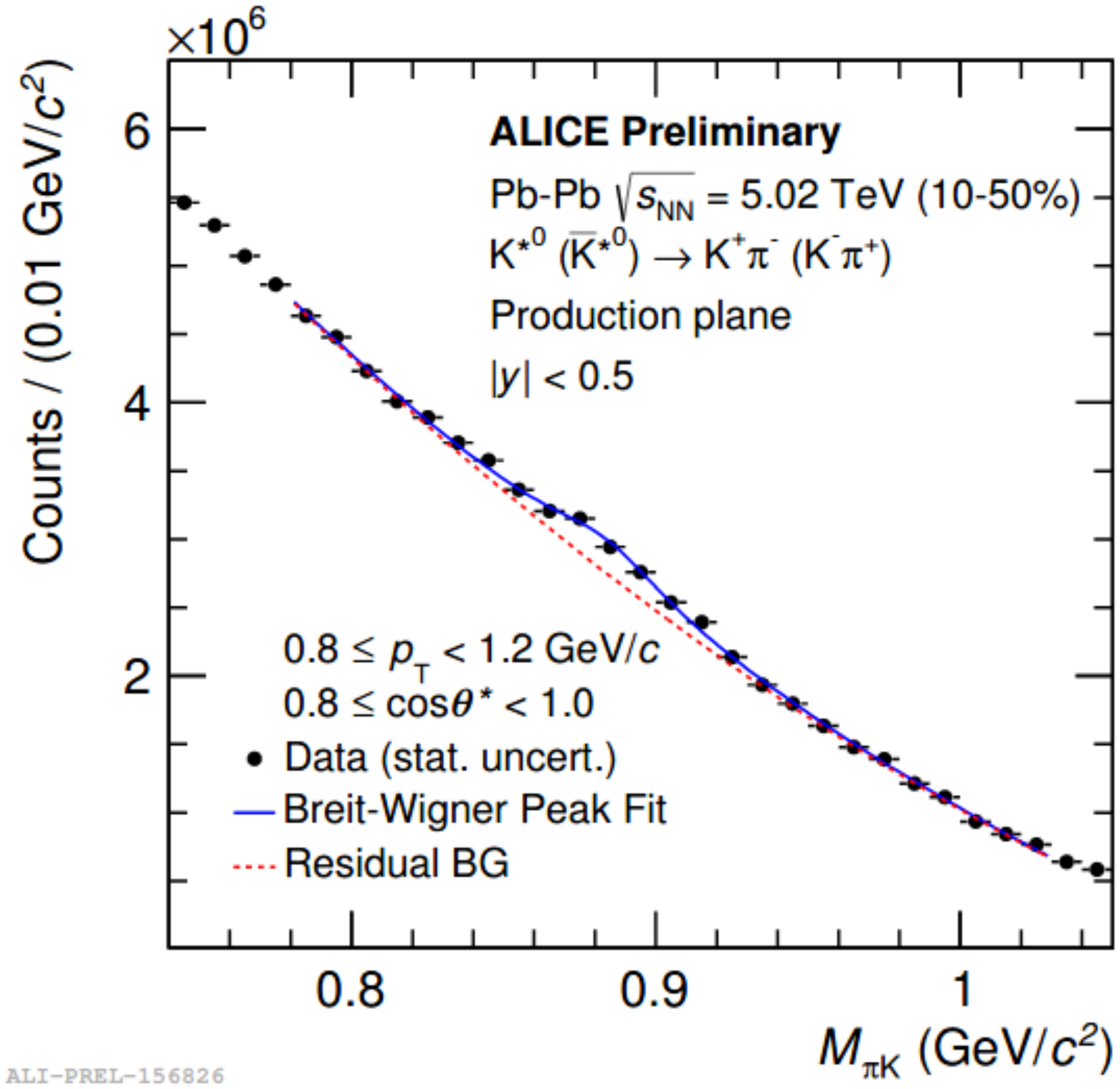}
\caption{Mixed event background subtracted invariant mass distribution of unlike charged $K\pi$ pairs in Pb--Pb collisions at \snn = 2.76 TeV, fitted with a Breit-Wigner function + 2$^{\mathrm{nd}}$ order polynomial function in $M_{K\pi}$. This distribution is obtained for 0.8 $<$ \pt $<$ 1.2 GeV/$c$ and 0.8 $<$ \ct $<$ 1.0, and the quantization axis is along the perpendicular direction to the event plane. This figure is taken from ~\cite{rsingh}}
\label{invmass} 
\vspace{-3mm}
\end{figure}

\begin{figure}
\begin{center}
\includegraphics[scale=0.5]{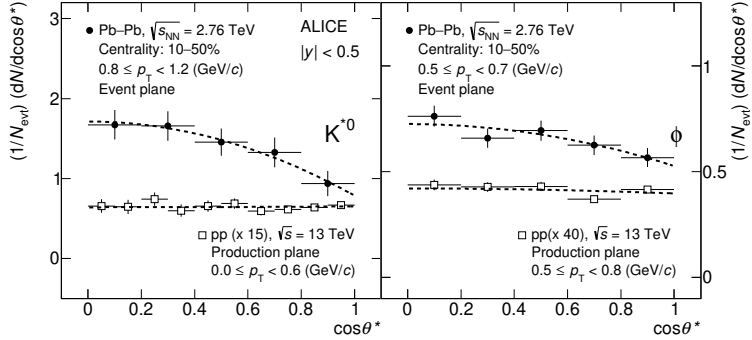}
\caption{Acceptance and efficiency corrected angular distribution of the decay daughter of \kst and \ph vector mesons in Pb--Pb collisions at \snn = 2.76 TeV and in pp collisions at $\sqrt{s}$ = 13 TeV. Measurements are carried out in $|y|$ $<$ 0.5. Each distribution is fitted with Eq.~(\ref{ang_dist}) to obtain \rh value. Statistical uncertainty on data points are represented by the vertical line.}
\label{ctdist}
\end{center}
\vspace{-3mm}
\end{figure}

Figure~\ref{invmass} shows a typical invariant mass distribution of unlike charge $K\pi$ pairs after mixed event combinatorial background subtraction in Pb--Pb collisions as a representative plot for \kst signal extraction. The invariant mass distribution is fitted with combination of 
a Breit-Wigner and a second order polynomial function. Area under the Breit-Wigner distribution is the measured \kst yield in a given \pt and \ct interval. The acceptance $\times$ efficiency corrected \kst and \ph meson yield as a function of \ct for selected \pt intervals in pp collisions at $\sqrt{s}$ = 13 TeV and in mid-central Pb--Pb collisions at \snn = 2.76 TeV are shown in Fig.~\ref{ctdist}. These distributions are fitted with Eq.~(\ref{ang_dist}) to obtain \rh values. The angular distributions for \kst and \ph vector mesons in Pb--Pb collisions at \snn = 2.76 TeV with respect to the normal of the event plane and in pp collision at $\sqrt{s}$ = 13 TeV with respect to the normal of the production plane are shown in Fig.~\ref{ctdist}.

\section{\label{results_pp}Spin alignment measurements in small collision system}
Spin alignment measurements of light flavour vector mesons have been carried out in helicity frame for $e^{+}e^{-}$ collisions, in order to understand the vector meson production mechanism. There are several models of vector meson production which predict the \rh values in the helicity frame. Spin counting based statistical model~\cite{statmodel1,statmodel2} assume that the fragmentation process produces extra quark--antiquark pairs with all helicity states being equally probable. In this model framework, a vector meson with helicity $\lambda$ = $\pm$1 will be produced if the spin of primary and secondary quarks are in parallel. If the spin of primary and secondary quarks are in antiparallel, then either a vector meson or a pseudo-scalar meson could be produced with probability 1 - $f$ and $f$. This production mechanism of vector meson leads to a \rh value of $\frac{1~-~f}{1~-~2f}$ which ranges between 0 to 0.5. In terms of the ratio of pseudoscalar to vector meson production ration ($P/V$) the \rh can also be expressed as \rh = $\frac{1}{2}~(1-(P/V))$. 
The QCD-inspired model, discussed in~\cite{qcd1} describe the fragmentation of soft gluons which are emitted from the fast primary quark. Soft gluons are fragmented into quark-antiquark pairs and at the end of the fragmentation chain the soft antiquark combine with the fast quark of same helicity to produce a vector meson. In this model framework produced vector mesons have \rh = 0. On the other hand, QCD based model described in~\cite{statmodel1} predicts the production of of vector mesons with $\lambda$ = 0. In this model framework vector mesons are produced through the channel q$\rightarrow$qV and the coupling of vector meson with quark is like a vector current coupling. In such case, helicity conservation of vector current ensures the vector meson production with 
$\lambda$ = 0 which corresponds to \rh=1. Similar prediction for \rh has also been predicted by other QCD based model~\cite{pqcd1,pqcd2,pqcd3} in which the hadron energy is equally shared by constituent quarks.
 %On the other hand, QCD based model describes in~\cite{statmodel1} predicts the production of of vector mesons with $\lambda$ = 0. In this model framework vector mesons are produced through the chanel q$\rightarrow$qV, with the vector meson coupling to the quark like a vector current. In such case, helicity conservation of vector current ensures the produced vector meson have $\lambda$ = 0 corresponding to \rh =1. Similar prediction for \rh has been also predicted by other pQCD based model~\cite{pqcd1,pqcd2,pqcd3} in which the hadron energy is equally shared by constituent quarks.\\

\begin{figure}
\begin{center}
\includegraphics[scale=0.5]{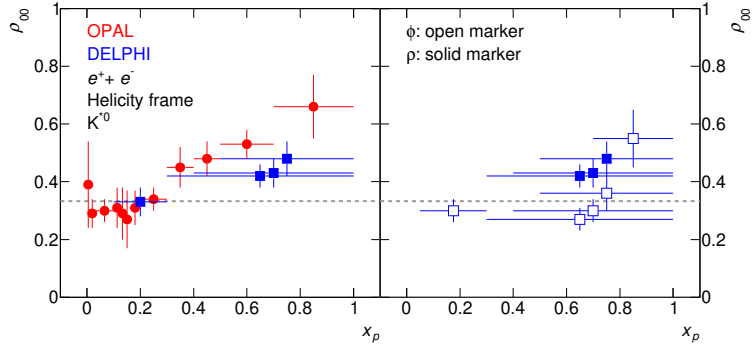}
\caption{Spin density matrix element \rh as a function of $x_{p}$ for \kst, \ph and $\rho$ meson in $e^{+}e^{-}$ collisions~\cite{opal1,opal2,delphi}. Measurements were carried out in the helicity frame. Uncertainties on data points are the quadrature sum of statistical and systematics uncertainties.}
\label{figlep}
\end{center}
\vspace{-3mm}
\end{figure}

The OPAL~\cite{opal1,opal2} and DELPHI experiments~\cite{delphi} have measured the spin density matrix element \rh in helicity frame. The \rh is larger than 1/3 for light flavour vector mesons \kst, \ph and $\rho$ in high $x_{p}$ ($\frac{p}{p_{beam}}$) region, whereas at low $x_{p}$ the measurements are consistent with 1/3. Measurements of \rh for $\rho$, \kst and \ph mesons, measured by DELPHI and OPAL collaboration are shown in Fig.~\ref{figlep}.
Measured \rh values for \kst and $\rho$ vector mesons are larger than 1/3 for $x_{p}$ $>$ 0.3, whereas DELPHI collaboration have observed \rh $>$ 1/3 for \ph meson for $x_{p}$ $>$ 0.7. Measurements from DELPHI collaboration are further confirmed by OPAL collaboration~\cite{opal2}, where they have also observed $\rh$ $>$ 1/3 for \ph meson for  $x_{E}$ ($E/E_{\mathrm{lab}}$) $>$ 0.7. This observation rules out the spin counting based statistical model~\cite{statmodel1,statmodel2} which gives the maximum \rh value of 0.5. However this can only happen when the probability of production of pseudo-scalar mesons containing the primary quark is zero. Observed spin alignment of vector mesons by DELPHI and OPAL collaboration  indicates the preferential production of vector meson in helicity zero state. This observation is consistent with the QCD based vector meson production mechanism, discussed in~\cite{statmodel1, pqcd1, pqcd2, pqcd3}. The OPAL and DELPHI experiments have also measured the off diagonal matrix element $\rho_{1-1}$ and they are found to be consistent with zero, ruling out the models~\cite{off1,off2,off3} which predict non-zero off-diagonal elements due to coherence phenomena. Within present uncertainties, no evidence has been observed for such effects.

\begin{figure}[hbtp]
\begin{center}
\includegraphics[scale=0.3]{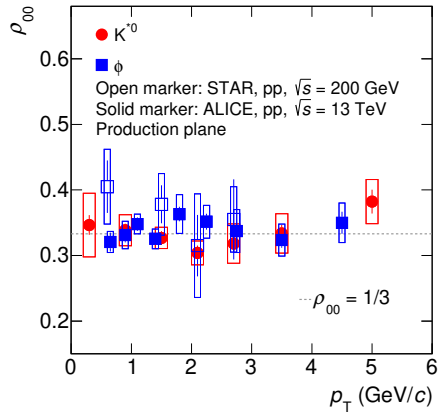}
\caption{Spin density matrix element \rh as a function of $p_{\mathrm{T}}$ for \kst and \ph vector mesons in pp collisions at $\sqrt{s}$ = 200 GeV~\cite{star200} and 13 TeV~\cite{alice}. Measurements were carried out in the production frame. Statistical and systematic uncertainties on data points are represented by bars and boxes, respectively.}
\label{figpp}
\end{center}
\vspace{-3mm}
\end{figure}

\begin{figure}[hbtp]
\begin{center}
\includegraphics[scale=0.4]{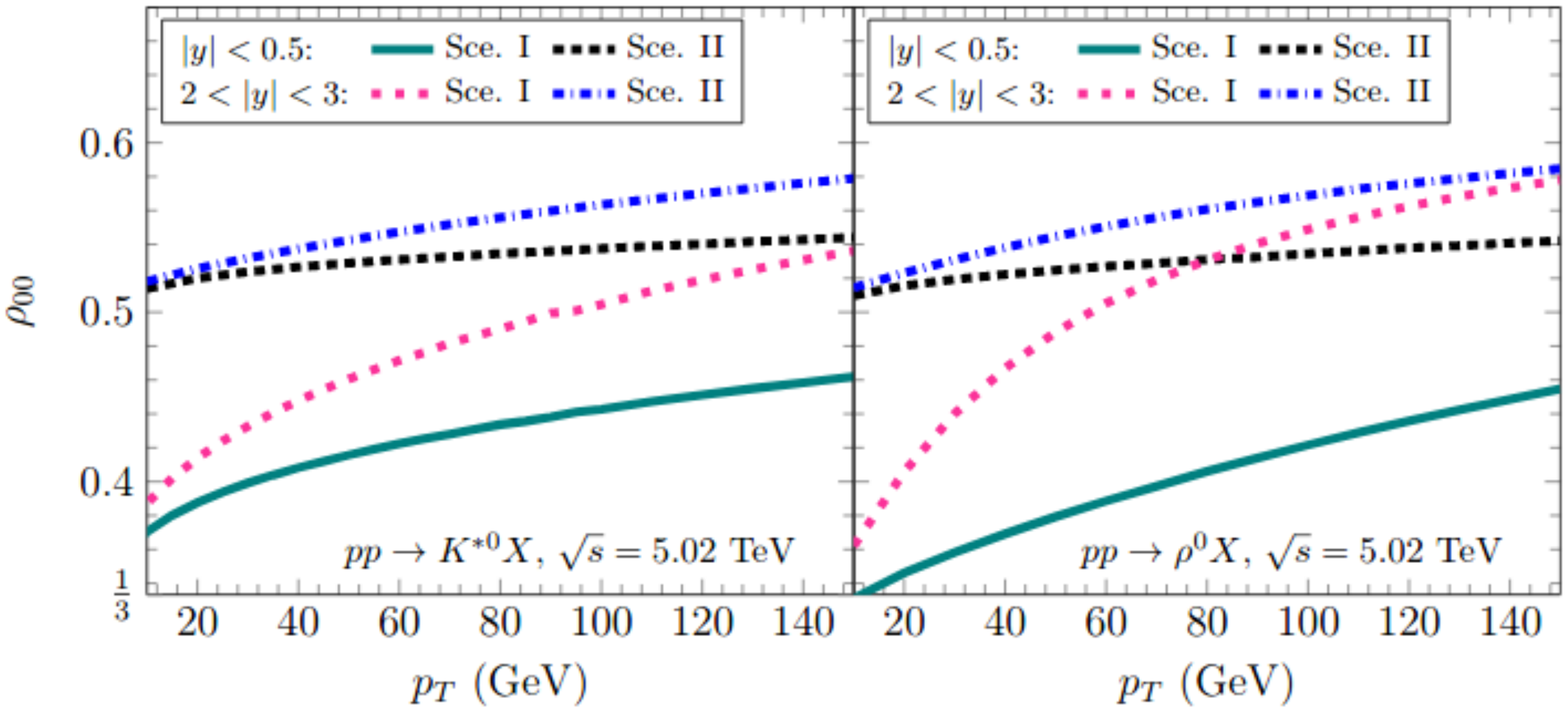}
\caption{Spin alignments of \kst and $\rho$ vector mesons in pp collisions at LHC energy $\sqrt{s}$ = 5.02 TeV in two rapidity regions as functions of \pt. The figure is taken from~\cite{theorypp}.}
\label{theorypp1}
\end{center}
\vspace{-3mm}
\end{figure}

\begin{figure}[hbtp]
\begin{center}
\includegraphics[scale=0.4]{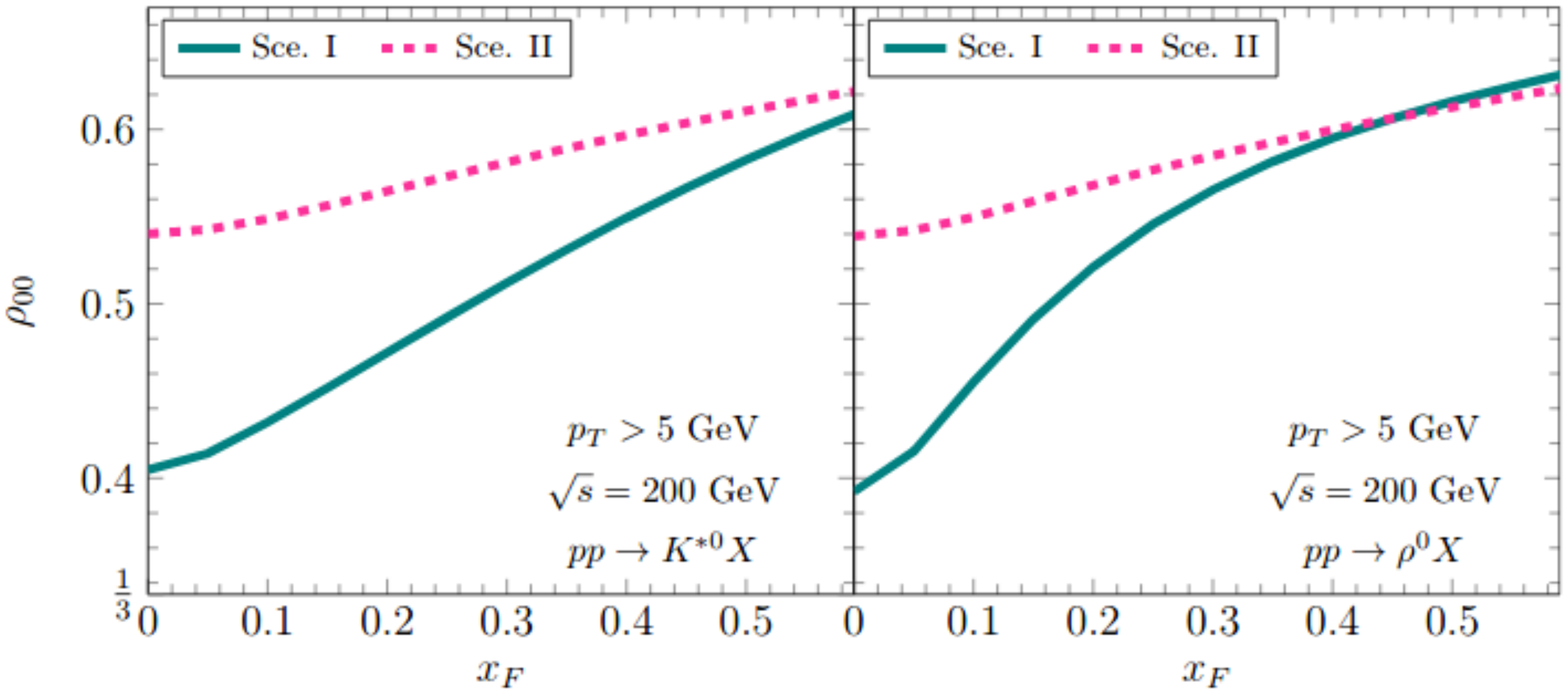}
\caption{Spin alignments of \kst and $\rho$ vector mesons in pp collisions at RHIC energy $\sqrt{s}$ = 200 GeV in two rapidity regions as functions of $x_{\mathrm{F}}$. The figure is taken from~\cite{theorypp}.}
\label{theorypp2}
\end{center}
\vspace{-3mm}
\end{figure}

Spin density matrix element \rh for \kst and $K^{*+}$ have also been studied in $kp$ and $nC$ interaction~\cite{kp1,kp2,kp3,kp4,kp5} with respect to the direction the perpendicular to production plane. The \rh values in these measurements are found to be significantly higher than the 1/3 which can be explained by a parton recombination model~\cite{ayala} and attributes the spin alignment of vector mesons via Thomas precession of the quark's spin in the recombination process of hadronization. Spin alignment measurements for \kst and \ph vector mesons with respect to the production plane have been also carried out in high energy pp collisions at the RHIC and the LHC. Figure~\ref{figpp} shows the measured \rh values for \kst and \ph vector mesons in pp collisions at $\sqrt{s}$ = 200 GeV~\cite{star200} and 13 TeV~\cite{alice}. Measured \rh values are consistent with 1/3 in the studied \pt region for \kst and \ph vector mesons at both RHIC and LHC energies. 
However, recent theoretical study in~\cite{theorypp} has predicted a significant spin alignment of vector mesons with respect to helicity frame in high energy pp collisions in fragmentation region (high \pt or high $x_{\mathrm{F}}$ = 2$p_{z}/\sqrt{s}$). In QCD, the spin alignment of vector meson produced in high energy reactions is determined by the spin-dependent fragmentation function $D_{1LL}$. In Ref.~\cite{theorypp} authors have used the measured \rh values for \kst from $e^{+}e^{-}$ collisions and estimated the \rh values for vector mesons in high energy pp collisions. They have found a significant spin alignment for \kst and $\rho$ vector mesons at high $x_{\mathrm{F}}$ and high \pt as shown in Fig.~\ref{theorypp1} and Fig.~\ref{theorypp2}. This theory prediction can be further tested in RHIC and LHC energies to understand the vector meson production mechanism at fragmentation region.

\section{\label{results_AA}Spin alignment measurements in heavy-ion collisions}
Spin alignment measurements have been recently carried out in heavy-ion collisions at RHIC and LHC energies in order to understand the spin-orbital angular momentum interaction. A large initial angular momentum and magnetic field is expected to be created perpendicular to the reaction plane in the initial stages of non-central (impact parameter $\sim$3--10 fm) heavy-ion collisions. The magnetic field is expected to be short-lived (a few fm/c), whereas the angular momentum is conserved, and its effect could present throughout the evolution of the system formed in the collision. In the presence of a large initial angular momentum produced quarks can be polarized due to the spin-orbital angular momentum interaction of QCD which further leads to the net polarization of vector mesons ($K^{*0}$ and $\phi$)~\cite{sam1,sam2,sam3,sam4,sam5}. Spin alignment measurement of vector mesons provide a unique opportunity to probe these initial conditions of heavy-ion collisions and can shed light on the presence of spin-orbital angular momentum interaction in heavy-ion collisions.

In theory there are some specific predictions from quark model for vector meson spin alignment and hydrodynamical calculation. In the the presence of a large angular momentum, the spin-orbit coupling of QCD could lead to a polarization of quarks followed by a net-polarization of vector mesons along the direction of the angular momentum~\cite{sam2,sam3,sam4,sam5}.  Quark model predicts~\cite{sam2,sam3,sam4,sam5} $\rho_{00}$ $<$ 1/3 if the vector meson is produced from the recombination of two polarized quarks. The recombination hadronization scenario polarized quark is expected to be observed at low \pt and in mid-rapidity. On the other hand quark polarization model predicts $\rho_{00}$ $>$ 1/3 if the hadronization of a polarized quarks proceeds via the fragmentation process.
Fragmentation hadronization scenario is expected to be observed at high \pt and in forward rapidity. In quark polarization model, polarization of quark is inversely proportional to square of its mass which suggests the spin alignment effect is larger for \kst than \ph due to their constituent quark composition. Recent theory calculation in~\cite{sam6} predicts the existence of coherence \ph meson field as a source of spin alignment of \ph meson which leads to $\rho_{00}$ $>$ 1/3 for the \ph meson. In quark model scenario, the initial large magnetic field may also affect the $\rho_{00}$~\cite{sam5}. For neutral vector mesons the magnetic field leads to $\rho_{00}$ $>$ 1/3, and for charged vector mesons it leads to $\rho_{00}$ $<$ 1/3. On the other hand, recent hydro-dynamical calculation~\cite{hydro} predicts \rh $<$ 1/3.
%$\propto~\frac{1}{3}~-~(\frac{\omega}{T})^{2}$, where $\omeg$ is the thermal vorticity and $T$ is the temperature.

\begin{figure}
\begin{center}
  \includegraphics[scale=0.4]{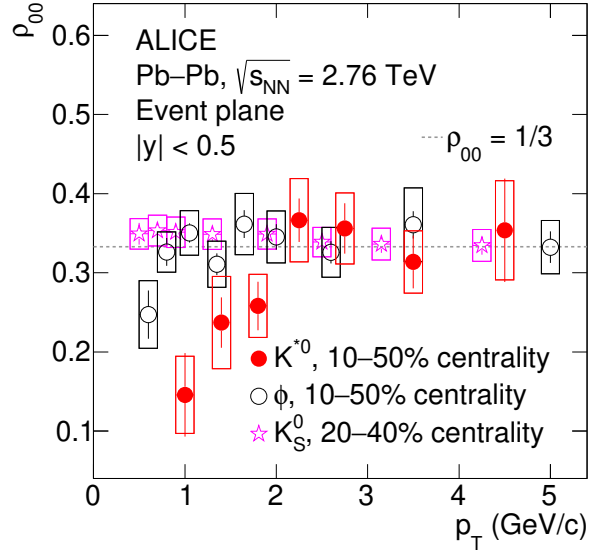}
\caption{The \pt dependence of \rh values for \kst, \ph, and \kzs mesons at $|y| <$ 0.5 in mid-central Pb--Pb collisions at \snn = 2.76 TeV~\cite{alice}. The statistical and systematic uncertainties are shown as bars and boxes, respectively.}
\label{Fig:momentum}
\end{center}
\vspace{-3mm}
\end{figure}

Figure \ref{Fig:momentum} shows \rh values as a function of \pt for \kst and \ph vector mesons, and \kzs meson at $|y| <$ 0.5 in Pb--Pb collisions at \snn = 2.76 TeV~\cite{alice} with respect to the normal to the event plane. Measured \rh values for the \kst and \ph with respect to the event plane (EP) in 10--50$\%$ Pb--Pb collisions at \snn = 2.76 TeV deviate from 1/3 at low \pta, whereas at high \pta, measurements are consistent with 1/3. For spin 0 hadron \kzs in  20--40$\%$ Pb--Pb collisions measured \rh values are consistent with 1/3 throughout the whole measured \pt interval as expected. Measured \rh values for vector mesons are consistent with 1/3 in control measurements such as pp collisions where the initial angular momentum is not expected. ALICE collaboration has also performed the spin alignment studies with respect to a quantization axis which is random in 3 dimension and the measured \rh values are consistent with 1/3 as expected~\cite{alice}.

ALICE collaboration has also measured the spin density matrix element \rh for vector mesons with respect to the normal direction of the production plane and the random event plane and these measurements are related to the measurements with respect to the normal direction of the event plane. The relation between measured \rh with respect to two different frames of references is
 \begin{equation}
\rh(\mathrm{A})-\frac{1}{3}=\left(\rh(\mathrm{B})-\frac{1}{3}\right)\left(\frac{1}{4}+\frac{3}{4}\cos{2\psi}\right),
\label{eq15} 
\end{equation}
where frame A is obtained by rotating frame B by angle $\psi$. Averaging over angle $\psi$ gives,
 \begin{equation}
\rh(\mathrm{A})-\frac{1}{3}=\left(\rh(\mathrm{B})-\frac{1}{3}\right)\left(\frac{1}{4}+\frac{3}{4}\langle\cos{2\psi}\rangle\right).
\label{eq16} 
\end{equation}
The transformation from the event plane (EP) to production plane (PP) is obtained by taking into account the elliptic flow of the vector meson which leads to
 \begin{equation}
\langle\cos{2\psi}\rangle=\frac{1}{2\pi}\int_{-\pi}^{\pi} \cos(2\psi)[1+2v_{2}\cos(2\psi)]d\psi=v_{2}.
\label{eq17} 
\end{equation}
Using Eq.~\ref{eq16} and Eq.~\ref{eq17}, analytical relation between EP and PP can be expressed as,
 \begin{equation}
\rh(\mathrm{PP})-\frac{1}{3}=\left(\rh(\mathrm{EP})-\frac{1}{3}\right)\left(\frac{1+3v_{2}}{4}\right).
\label{eq18} 
\end{equation}
In order to verify~\ref{eq18} a toy model simulation with PYTHIA (version 8.2) event generator~\cite{spin_pythia} is carried out. PYTHIA does not have any azimuthal anisotropy and spin alignment. For this study we have taken event plane angle as zero, which corresponds impact parameter along $x$-axis. In order to find the relations between different frames, $v_{2}$ (0.15 $\pm$ 0.06, value expected for hadrons with mass similar to \kst in \pb collisions at \snn = 2.76 TeV~\cite{Abelev:2014pua}) is introduced to \kst by appropriate rotation of its momentum in azimuthal plane. The modified angle in azimuthal plane is calculated by solving the following equation

\begin{equation}
\phi_{0}=\phi+v_{2}\sin2\phi,
\label{eq30}
\end{equation}
where $\phi_{0}$ is azimuthal angle of \kst in absence of $v_{2}$ and for a given value of $v_{2}$, $\phi_{0}$ transforms to $\phi$. Then spin alignment through \rh same as measured in data is introduced with respect to event plane by rotating the momentum of decay daughters in \kst rest frame by solving
\begin{equation}
\cos{\theta^{*}_{0}} = 3/2 \times [(1-\rho_{00}) \cos{\theta^{*}} + 1/3(3\rho_{00}-1)\cos^{3}\theta^{*}].
\label{eq31} 
\end{equation}
\begin{figure}
\begin{center}
\includegraphics[scale=0.4]{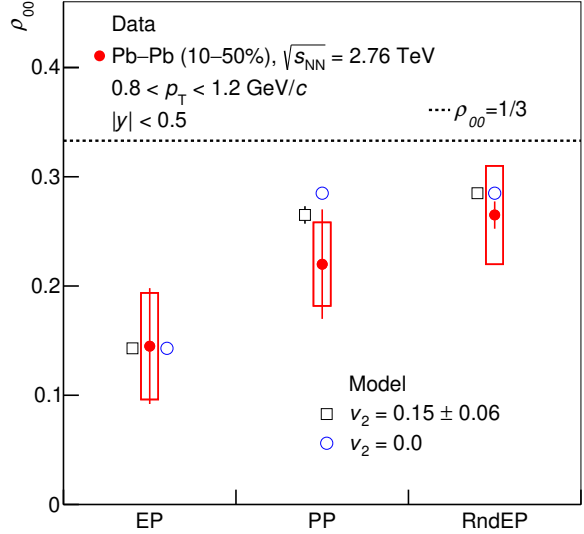}
\caption{\rh values from data in 10--50\% Pb--Pb collisions at 0.8 $<$ \pt $<$ 1.2 GeV/$c$ with respect to various planes~\cite{alice} compared with expectations from model simulations with and without added elliptic flow ($v_{2}$). The statistical and systematic uncertainties are shown as bars and boxes, respectively. This figure has been taken from~\cite{alice}.} 
\label{toymodel}
\end{center}
\vspace{-3mm}
\end{figure}  
Here $\theta^{*}_{0}$ is the angle made by the decay daughter of \kst with the quantization axis in absence of spin alignment. $\theta^{*}_{0}$ transforms to $\theta^{*}$ to introduce a given input value of \rh. In this study we assume that the $\phi^{*}$ remain fixed during the rotation. With these modifications, calculations as in the experimental data are carried out. The results are shown in Fig.~\ref{toymodel} for two cases, with and without $v_{2}$. The result corresponding to event plane is correctly retrieved in the model. The model results for $v_{2}$ = 0, are same for production plane and random event plane. However with $v_{2}$ = 0.15, the \rh(PP) value is lower and closer to data for PP. The toy model reproduces the hierarchy observed in the \rh values for various planes as observed in data. RndEP is defined by randomizing the event plane angle in the azimuthal plane ($xy$ plane). The quantization axis is obtained by taking cross product of event plane vector and $\mathrm{z}$-axis (beam axis). Hence, the quantization axis is always in the $xy$ plane, and a residual effect is present due to use of a common $\mathrm{z}$-axis. Decay daughters of polarized vector mesons could have angular correlation with respect to the $\mathrm{z}$-axis. This residual effect is due to the angular distribution of decay daughters, which follow a oblate or prolate shape, which is rotationally symmetric around the quantization axis (when the off-diagonal terms in the spin density matrix are zero). Rotating such a shape around the $\mathrm{z}$-axis does not lead to a uniform decay angular distribution. This effect results in small deviation of \rh values from 1/3 for RndEP. This deviation of \rh from 1/3 for RndEP has been also discussed in Ref~\cite{sa_res}. The physical picture is that spin alignment with respect to the event plane is coupled to that in the production plane through the elliptic flow of the system. The \rh(RndEP) is lower than 1/3 as the quantization axis is always perpendicular to the beam axis, resulting in a residual effect. This residual effect can be removed by choosing the quantization axis in a random direction in 3 dimensions for each event. If the quantization axis is random in 3 dimension, then the residual effect is not present and the \rh value is consistent with 1/3.
\begin{figure}
  \begin{center}
    \includegraphics[scale=0.5]{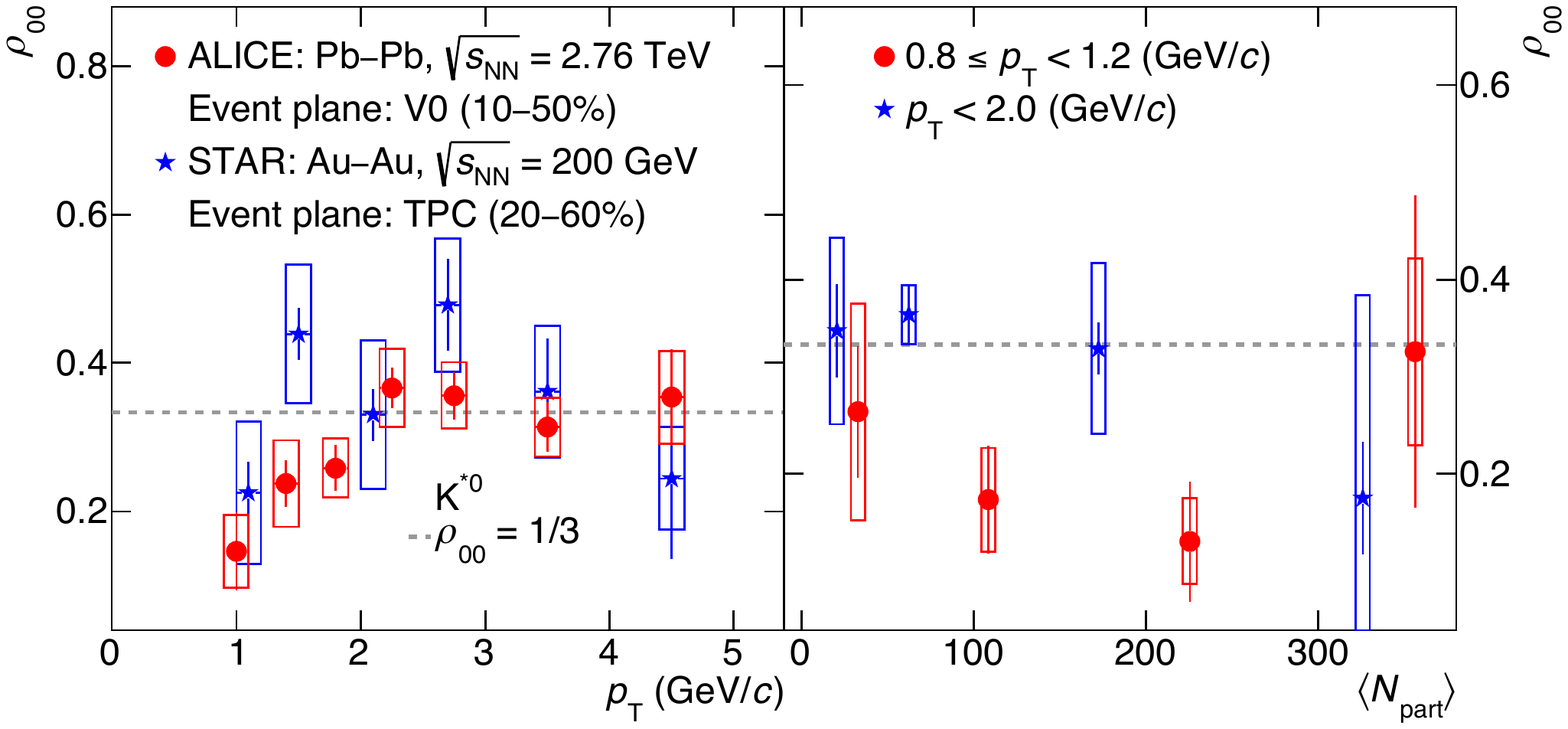}
\caption{Left panel: \rh vs. \pt for \kst meson at $|y| <$ 0.5 in mid-central Pb--Pb and Au--Au collisions at \snn = 2.76 TeV~\cite{alice} and 200 GeV~\cite{star2007}. The statistical and systematic uncertainties are shown as bars and boxes, respectively.}
\label{alicestarcomp_ptcent}
\end{center}
\end{figure}
On the other hand, preliminary spin alignment measurements of \ph meson at RHIC energies~\cite{starphi} shows \rh $>$ 1/3 in intermediate \pt (above 1.2 GeV/$c$) which is in contrast with the ALICE measurement that shows a deviation (\rh $<$ 1/3) at low \pt below 0.7 GeV/$c$ where the measurements at RHIC energies do not exist till date. Figure~\ref{staralicept} shows the comparison of \pt dependence of \rh for \ph meson between Pb--Pb and Au--Au collisions at \snn = 2.76 TeV and 200 GeV, respectively. 
\begin{figure}
\begin{center}
  \includegraphics[scale=0.4]{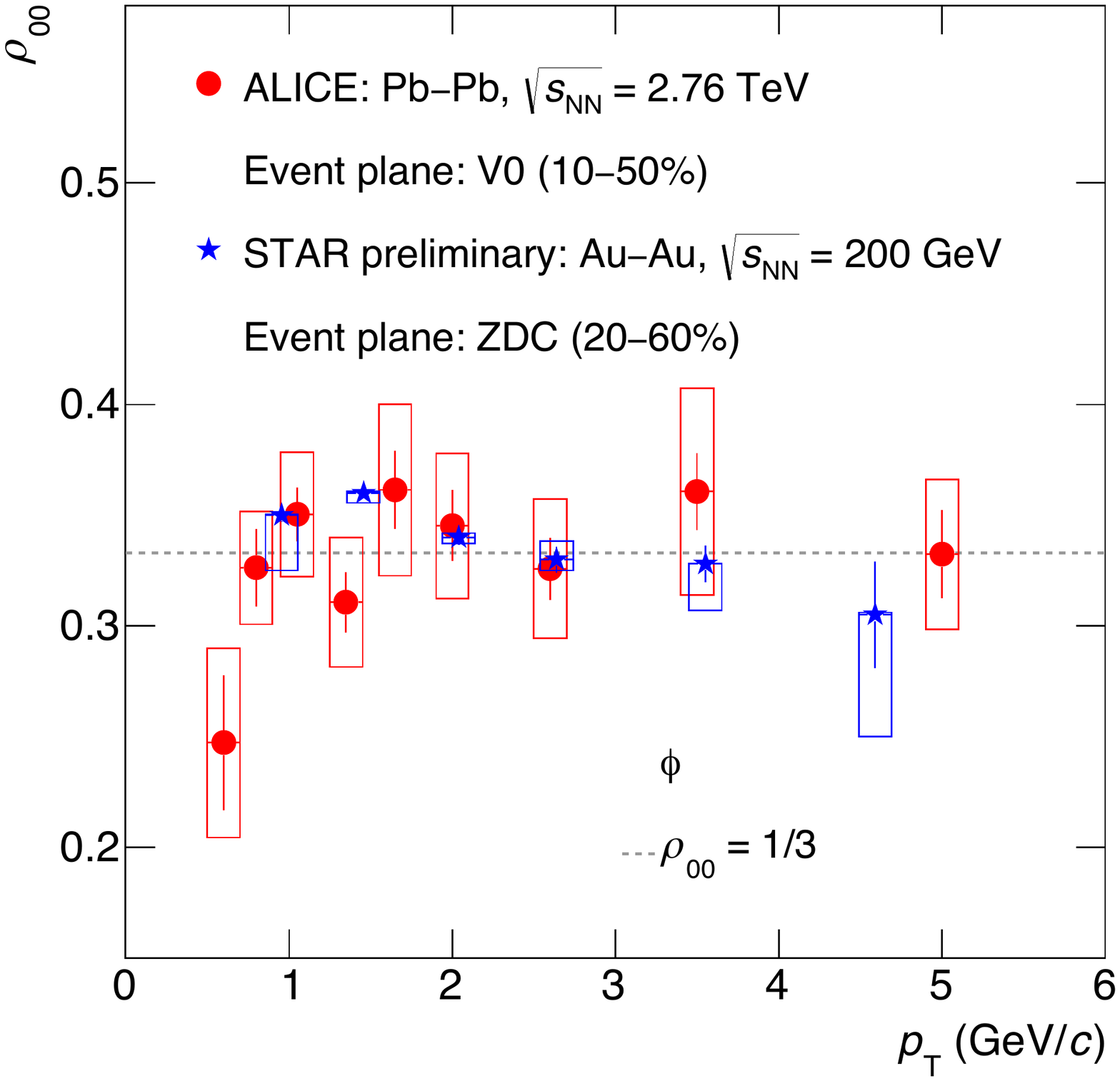}
\caption{Spin density matrix element \rh as a function of \pt for \ph meson in mid-central Pb--Pb and Au--Au collisions at \snn = 2.76 TeV~\cite{alice} and 200 GeV~\cite{starphi}. The statistical and systematic uncertainties are shown as bars and boxes, respectively.}
\label{staralicept}
\end{center}
\end{figure}

ALICE collaboration has also observed a clear centrality dependence of spin alignment for \kst and \ph meson at low \pta~\cite{alice}. Deviation of \rh from 1/3 is maximum at mid central collisions, whereas in central and peripheral collisions \rh values are consistent or closed to 1/3. The observed centrality dependence of spin alignment for \kst and \ph vector mesons are consistent with the impact parameter dependence of initial angular momentum. 
%Recent spin alignment measurements of \kst vector meson by STAR collaboration~\cite{starsubhash} also show similar \pt and centrality dependence of measured \rh values. 
Figure~\ref{alicestarcomp_ptcent} shows comparison of \pt and centrality dependence of \rh for \kst in Pb--Pb collisions at \snn = 2.76 TeV and Au--Au collisions at \snn = 200 GeV measured by ALICE and STAR collaboration, respectively.   

\begin{figure}
\begin{center}
  \includegraphics[scale=0.6]{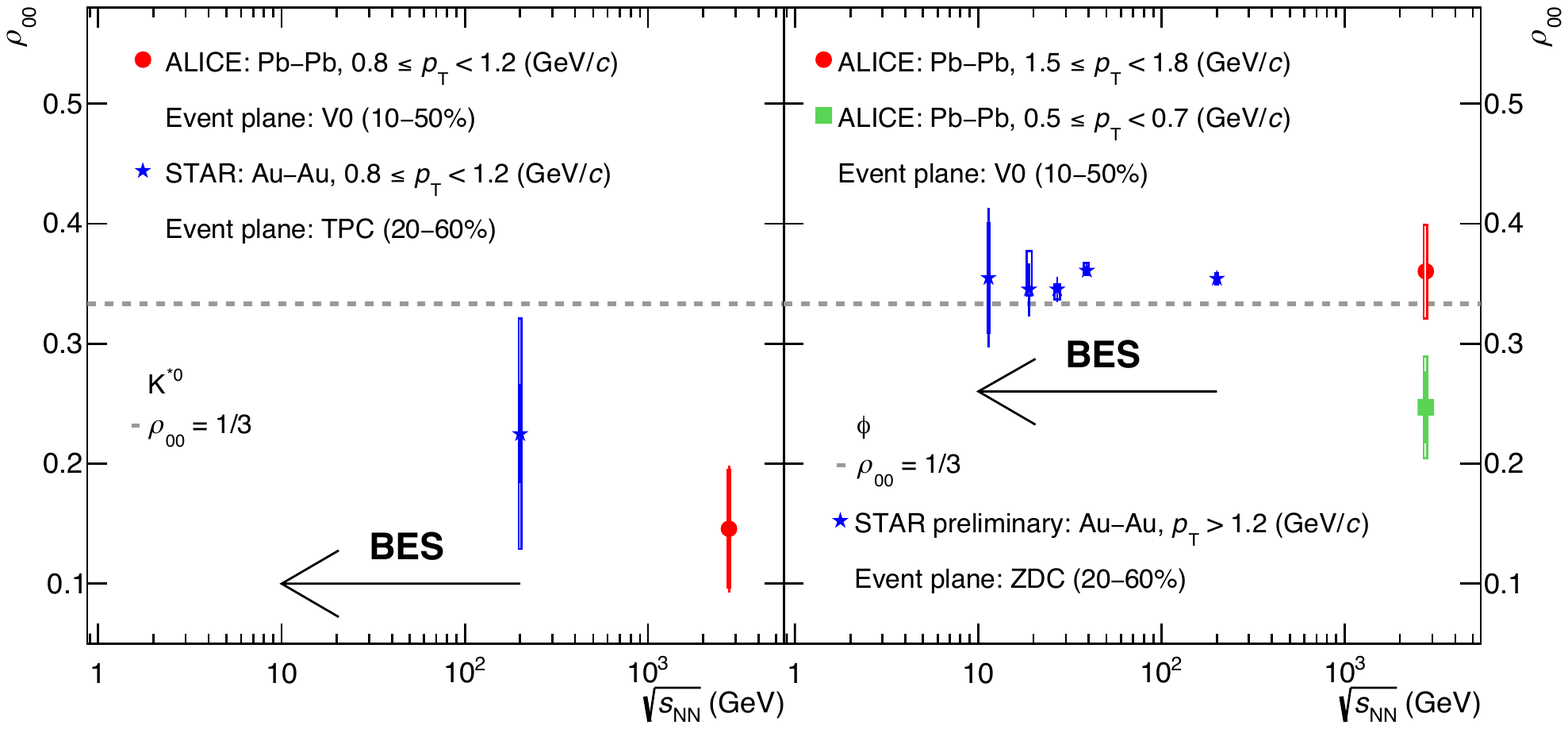}
\caption{Beam energy dependence of \rh for \kst~\cite{star2007,alice} and \ph~\cite{starphi,alice} vector mesons. Statistical and systematic uncertainties are shown by bars and boxes, respectively. BES represents the region to be explored by the beam energy scan program at RHIC.}
\label{figbes}
\end{center}
\end{figure}

The observed \pt dependence of \rh for \kst vector meson in heavy-ion collisions at both LHC and RHIC energies and for \ph meson at LHC energy are qualitatively consistent with the prediction from the quark recombination model of polarized quarks~\cite{sam2,sam3,sam4,sam5} which attributes the spin alignment to polarization of quarks in the presence of large initial angular momentum in non-central heavy-ion collisions and a subsequent hadronization by the process of recombination. However, \rh for \ph mesons are larger than 1/3 in mid-central Au--Au collisions at \snn = 200 GeV~\cite{starphi}. The \rh $>$ 1/3 for \ph mesons can not be explained by naive quark recombination and fragmentation model~\cite{sam2} but may come from the electric part of the \ph meson field~\cite{sam6}. The centrality or $\langle N_{\mathrm {part}} \rangle$ dependence of \rh is qualitatively consistent with the variation of initial angular momentum with impact parameter in heavy-ion collisions~\cite{sa1}. In recombination hadronization scenario of polarized quarks, the \rh of vector mesons are related to quark polarization as $\rh = \frac{1-P_{\mathrm{q}}P_{\bar {\mathrm{q}}}}{3+P_{\mathrm{q}}P_{\bar {\mathrm{q}}}}$ where $P_{\mathrm{q}}$ and $P_{\bar {\mathrm{q}}}$ are the polarization of quark and anti-quark in presence of the initial angular momentum~\cite{sam2}. Assuming the same quark polarization for light and strange quark and antiquark, measured \rh values can be used to estimate the order of quark polarization. %Figure~\ref{quarkpol} shows the estimated quark polarization as a function of \snn using \rh measurements for \kst at RHIC~\cite{starsubhash} and LHC energies~\cite{alice}. For ALICE data \rh measurements from a low \pt bin is used as the result for integrated \pt is not available, while the \rh value in integrated \pt range (same as STAR collaboration) would be slightly lower hence would slightly decrease the quark polarization value. 
Estimated quark polarization from quark recombination model is of the order of $O$(0.1)~\cite{alice} and with the present uncertainties no significant energy dependence of quark polarization is observed.

%\begin{figure}[hbtp]
%\begin{center}
%  \includegraphics[scale=0.4]{figure/quark_pol.eps}
%\caption{Quark polarization values as a function of \snn estimated from the \rh measurements of %\kst~\cite{starsubhash,alice}. Uncertainties on quark polarization are obtained from the total %uncertainties (quadrature sun of statistical and systematic uncertainties) on \rh values.}
%\label{quarkpol}
%\end{center}
%\vspace{-3mm}
%\end{figure}
%%
Although the quark recombination model~\cite{sam2,sam3,sam4,sam5} qualitatively describe the observed \pt dependence of \rh for \kst vector meson in heavy-ion collisions at both LHC and RHIC energies and for \ph meson at LHC energy but quantitatively the experimental results are surprisingly high compared to the model calculation in context of $\Lambda$ polarization. While no quantitative theoretical calculation for vector meson spin alignment at LHC energies exists, the expected order of magnitude can be estimated and the measurements for vector mesons and hyperons can be related in a model dependent way. The polarization of $\Lambda$ baryons (spin = 1/2) is about $O$ (1\%) at lower RHIC energies~\cite{star_lambda1,star_lambda2} and decreases with energy to reach $O$ (0.1\%) at LHC energy~\cite{alice-gp}. In the quark recombination model, $\Lambda$ polarization linearly depends on quark polarization~\cite{sam2} (p${_\mathrm{q}}$), whereas \rh depends on the square of quark polarization~\cite{sam3}. With these assumptions and taking the input of p${_\mathrm{q}}$ from $\Lambda$ polarization measurements at LHC energies~\cite{alice-gp}, one would expect spin alignment effect is very tiny and \rh is closed to 1/3. Therefore, the observed spin alignment of vector meson is surprisingly large in context of $\Lambda$ polarization. Using a thermal and non-relativistic approach as discussed in~\cite{hydro}, vorticity ($\omega$) and temperature ($T$) are related to hyperon polarization ($P_{\Lambda}$) and \rh as $P_{\Lambda}$ $\simeq$ $\frac{1}{4}\frac{\omega}{T}$ and \rh $\simeq$ $\frac{1}{3}\left(1-\frac{\left( \omega/T\right)^{2}}{3}\right)$, respectively. In this model framework, estimated spin alignment for vector mesons from $\Lambda$ polarization measurements are also very tiny and \rh $\simeq$ 1/3.

However, for spin 1/2 particles the angular distribution of the decay daughters is an odd function and hence depends on the sign of the angular momentum direction. For spin 1 particles the angular distribution of the decay daughters is an even function and only depends on the strength of the angular momentum not on the sign. Unlike vector mesons, hyperons also have a large decay contribution. The \rh value may also depend on the details of the transfer of the quark polarization to the hadrons during hadronization. In addition, re-scattering, regeneration, lifetime and mass of the hadron, and interaction of the hadron with the medium may also affect the magnitude of \rha. So far, the large magnitude of global spin alignment of \kst and \ph vector mesons compared to $\Lambda$ polarization, observed in high energy heavy-ion collisions at RHIC and LHC energies cannot be explained by conventional effects such as vorticity field, electromagnetic field and recombination hadronization of polarized quarks. Recent theory development discussed in Ref.~\cite{sam6,improvedquark} suggests a  large positive deviation of \rh for \ph mesons from 1/3 may come from the electric part of the \ph meson field. Existence of \ph meson field in high energy heavy-ion collisions may explain the significantly large spin alignment of \ph vector mesons at RHIC energies. Vector meson field formalism is difficult to develop for \kst meson because of the unequal masses of s and d quarks. Due to the unequal masses of constituent quarks, decoupling of the contribution from the vorticity, electric field, magnetic field and vector meson field is difficult for \kst. In addition, due to the short lifetime of \kst meson interaction of \kst with the surrounding matter is much stronger than that of the \ph meson and this may also affect the \rh values of \kst~\cite{dshen}. Recent study in Ref.~\cite{local} shows that the deviation of \rh from 1/3 may also comes from the local spin alignment of vector mesons. In this model framework, the large negative deviation of \rh from 1/3 (\rh $<$ 1/3) is not solely coming from the global spin alignment but have also contribution from the local polarization of quarks and anti-quarks which are locally polarized due to the anisotropic expansion of the medium. However, the contribution from the local quark polarization in \rh is yet not constrained from the experiments as it requires the measurements for off diagonal spin density matrix elements. Till now, no quantitative theory explanation for the measured spin alignment of vector meson at RHIC and LHC energies exists. Therefore these measurements will trigger further theoretical works to understand the results.

The measured $\Lambda$ polarization value at RHIC~\cite{star_lambda1,star_lambda2} and LHC~\cite{alice-gp} energies are found to be decreases with \snn. In quark model, non-relativistic hydrodynamics and \ph meson field approach, estimated spin alignment of vector mesons from $\Lambda$ polarization measurements reduces with \snn. However, within the present statistical uncertainties no such energy dependence of vector meson spin alignment is observed, as shown in Fig.~\ref{figbes}. In future, high statistics BES II data at RHIC energies and in Pb--Pb collisions at 5.02 TeV data at LHC will help to look at the \snn dependence of vector meson spin alignment with precise measurements. 

\section{\label{summary}Summary and outlook}
The spin alignment measurements have been studied in $e^{+}e^{-}$ and hadron collisions for decades to understand the vector meson production mechanism. Measurements of spin density matrix element \rh with respect to the helicity frame in $e^{+}e^{-}$ collisions~\cite{opal1,opal2,delphi} shows an evidence of spin alignment of vector mesons produced in high $x_{p}$ region. The \rh values are found to be larger than the 1/3 for \ksta, \ph and $\rho$ mesons. This observation suggests that the fragmentation of an unpolarized quark led to vector meson with a larger probability at the helicity zero state. This observations are consistent with the QCD based model~\cite{statmodel1} in which vector meson production is arising from the helicity conserving process q$\rightarrow$qV. Other QCD models~\cite{pqcd1,pqcd2,pqcd3} based on the equal distribution of hadron energy in constitutent quarks also prefer the production of vector meson with helicity state zero.  On the other hand measurements of off diagonal spin density matrix elements are consistent with zero ruling out coherence models~\cite{off1,off2,off3} which predicted non zero off diagonal elements. The spin alignment measurements for \kst and $K^{*+}$ vector mesons have also been studied in $kp$ and $nC$ interaction~\cite{kp1,kp2,kp3,kp4,kp5} with respect to the direction perpendicular to the production plane. In these experiments, the measured \rh values are found to be larger than the 1/3 and can be explained by a parton recombination model~\cite{ayala} which attributes to the spin alignment of vector mesons via Thomas precession of the quark's spins in the recombination process of hadronization. Measurements for spin density matrix element \rh have been further extended in pp collisions at \s = 200 GeV~\cite{star200} and 13 TeV~\cite{alice} with respect to the production plane. Measurements are found to be consistent with 1/3 in measured \pt region (\pt $<$ 5 GeV/$c$). In future, using high statistics pp collisions data at RHIC and LHC energies, these measurements can be carried out at high $x_{p}$ region where recent theory study in Ref.~\cite{theorypp} predicts a significant spin alignment for vector mesons.

Given the importance of spin-orbit interactions in several fields of physics, it is imperative to look for its possible effect on particles with spin in a system with high orbital angular momentum. Knowing the presence of the large angular momentum created in non-central heavy-ion collisions, it is important to study the spin alignment of vector mesons to understand the spin orbital angular momentum interaction in QCD matter. Recent measurements of \rh for \kst and \ph vector mesons in Pb--Pb collisions at the LHC and in Au--Au collisions at the RHIC reported significant spin alignment. The ALICE experiment~\cite{alice} has observed a significant spin alignment effect at a level of 3$\sigma$ for \kst and 2$\sigma$ for \ph in mid-central Pb--Pb collisions at \snn = 2.76 TeV due to the spin orbital angular momentum interaction. The spin density matrix element \rh is found to be lower than 1/3 at \pt $<$ 2 GeV/$c$, whereas at high \pt the measurements are consistent with 1/3. In order to validate the measurements, various control measurements such as measurements of \rh for vector mesons with respect to the random event plane, measurements of \rh for \kzs in mid-central heavy-ion collisions, and spin alignment measurements of vector mesons in pp collisions are carried out. In all control measurements measured \rh values are consistent with 1/3. The observed \pt dependence of measured \rh values for vector mesons in mid-central Pb--Pb collisions can be attributed as a presence of a large initial angular momentum in non-central heavy-ion collisions, which leads to quark polarization via spin-orbit coupling, subsequently transferred to hadronic degrees of freedom by hadronization via recombination. 
%Preliminary results from the STAR collaboration have also shown similar \pt dependence of measured \rh value for \ksta. 
However, the preliminary spin alignment measurements for \ph meson at RHIC energies show \rh $>$ 1/3 at intermediate \pta. The \rh $>$ 1/3 for \ph mesons can not be explained by naive quark recombination and fragmentation model but can be explained due to the presence of \ph meson field in heavy-ion collisions. The spin alignment effect of vector mesons is maximum in mid-central Pb--Pb collisions, whereas in central and peripheral collisions the \rh values are consistent with 1/3. The centrality dependence of spin alignment effect of vector meson is consistent with the impact parameter dependence of initial angular momentum. Although a significant spin alignment effect of vector mesons have been observed but the measured spin alignment of vector mesons is surprisingly large compared to the polarization measured for $\Lambda$ hyperons. Possible reason behind the large difference between $\Lambda$ polarization and vector meson spin alignment may include the transfer of the quark polarization to the hadrons (baryon vs. meson), details of the hadronization mechanism (recombination vs. fragmentation), re-scattering, regeneration, and possibly the lifetime and mass of the relevant hadron. Moreover, the vector mesons are predominantly directly produced whereas the hyperons have large contributions from resonance decays. So far, there are no quantitative theory explanation for the large \rh values of \kst and \ph vector mesons observed at RHIC and LHC energies and these measurements will trigger further theoretical works in order to study which effects could make such a huge difference between $\Lambda$ and vector meson polarization. However, recent theory developments which incorporate vector meson field and local polarization may explain the large spin alignment of vector mesons.

In future, high statistics data at the RHIC and LHC energies will provide precise measurements and can be useful to understand the beam energy dependence of \rha. High statistics data may allow the possibility of assessing the effect of initial magnetic field produced in heavy-ion collisions by measuring the difference in the spin alignment of $K^{*\pm}$ and $K^{*0}$ as the magnetic moment of $K^{*\pm}$ is $\sim$7 times larger than the \ksta. Measurements of the off diagonal spin density matrix elements and azimuthal angle (vector meson azimuthal angle with respect to the reaction plane angle) dependency of \rh can be used to disentangle the local spin alignment from the global spin alignment. In addition, extension of spin alignment study at high \pt in pp collisions will be useful to constrain the spin dependent fragmentation function.

\section*{Acknowledgments}
B.M. was supported in part J C Bose Fellowship from Department of Science of Technology, Government of India and Research in Basic Sciences project from Department of Atomic Energy, Government of India.
SS is supported by the Strategic Priority Research Program of Chinese Academy of Sciences (Grant XDB34000000).

\end{document}